\documentclass[prb,twocolumn,showpacs,amsmath,letter]{revtex4}
\usepackage{graphicx}

\newcommand{\Journal}[4]{#1 \textbf{#2}, #3 (#4)}

\begin{document}

\title{Effect of spin diffusion on spin torque in magnetic nanopillars}

\author{Sergei Urazhdin}
\author{Scott Button}
\affiliation{Department of Physics, West Virginia University, Morgantown, WV 26506}


\begin{abstract}

We present systematic magnetoelectronic measurements of magnetic nanopillars with different structures of polarizing magnetic layers. The magnetic reversal at small magnetic field, the onset of magnetic dynamics at larger field, and the magnetoresistance exhibit a significant dependence on the type of the polarizing layer. We performed detailed quantitative modeling showing that the differences can be explained by the effects of spin-dependent electron diffusion.
\end{abstract}

\maketitle

According to the spin torque (ST) model~\cite{slonczewski}, current-induced magnetic switching (CIMS) in magnetic multilayers is caused by angular momentum transfer from the conduction electrons to the magnetic layers. ST is believed to occur within atomic distances from the magnetic interfaces.  Nevertheless, theories have shown that electron diffusion in the layers has an important effect on ST.~\cite{kovalev,zhanglevyfert,shpiro} As a simple example, an electron scattering between two ferromagnets transfers angular momentum upon each reflection. However, this transfer is not necessarily associated with a net charge current $I$. Therefore, efficient utilization of electron scattering can result in reduced $I$ required to manipulate magnetic devices with ST. In a more subtle manifestation, spin-dependent electron diffusion causes an asymmetry between the ST in antiparallel (AP) and parallel (P) configurations of the magnetic layers.~\cite{slonczewski2002} In an extreme case of such asymmetry, ST can change direction, resulting in anomalous current-induced behaviors.~\cite{fertCoPy}

Despite extensive theoretical work, few experiments addressed the effects of diffusion on ST.~\cite{fertCoPy,iswvsmr,CoCo,mustafa,diffusive}  The main difficulty stems from the limited knowledge about the transport properties of individual layers in magnetic nanostructures. Different deposition and measurement techniques yielded significantly different values.~\cite{bassreview} On the other hand, both the Magnetoresistance (MR) and CIMS depend on the same spin-dependent transport properties. Therefore, simultaneous measurements of MR and CIMS, and their analysis within the same theoretical framework can lead to better understanding of the electron diffusion and its effect on ST.

\begin{figure}
\includegraphics[scale=0.85]{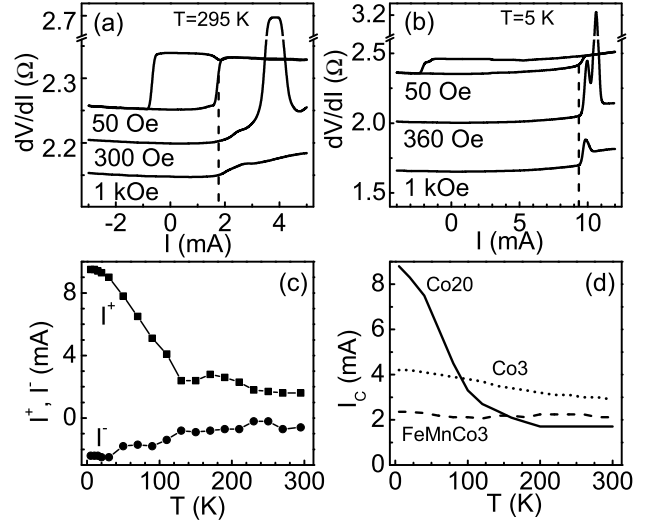}
\caption{\label{fig1} (a) $dV/dI$ {\it vs} $I$ at labeled $H$ and $T=295$~K. Curves are offset for clarity. (b)
same as (a), at $T=5$~K. (c) $I^+$, $I^-$ {\it vs} $T$ for a $Co20$ sample. (d) $I_C$ {\it vs} $T$ measured at $H=500$~Oe for the three types of samples as labeled.}
\end{figure}

We report systematic measurements of MR and CIMS in nanopillar spin valves F$_1$/N/F$_2$ with identical free  layers F$_2$=Py(5), Py=Ni$_{80}$Fe$_{20}$, and different polarizers F$_1$ incorporating Co. Thicknesses are given in nm. Large spin diffusion length $l_{sf,Co}$ makes Co ideal for studying the effects of diffusion. We used F$_1$=Co(20) in samples labeled $Co20$.  To separate the contributions of the Co interfaces and its bulk, we tested samples with F$_1$=Co(3), labeled $Co3$, in which the scattering in the bulk of Co(3) was negligible. To eliminate the spin diffusion in the sample contacts, we inserted a strongly spin flipping bilayer Fe$_{50}$Mn$_{50}$(1)/Cu(1) between Co(3) and the bottom contact in samples labeled $FeMnCo3$.

The multilayers Cu(50)/F$_1$/Cu(10)/F$_2$/Cu(200) were deposited at room temperature $295$~K (RT) by magnetron sputtering at base pressure of $5\times 10^{-9}$~Torr, in $5$~mTorr of purified Ar.  F$_2$ and part of the Cu(10) spacer were patterned into an elliptical nanopillar with approximate dimensions $130 \times 60$~nm. We measured $dV/dI$ with four-probes and lock-in detection. Positive $I$ flowed from $F_1$ to $F_2$. Magnetic field $H$ was in the film plane and along the nanopillar easy axis. At least three nanopillars of each  type were tested with similar results.

Figs.~\ref{fig1}(a),(b) show $dV/dI$ {\it vs} $I$ for a $Co20$ sample, acquired at RT and $5$~K, respectively. The data at small $H=50$~Oe are characterized by hysteretic jumps to the P state with low resistance $R_P$ at $I^-<0$, and to the AP state with high resistance $R_{AP}$ at $I^+>0$. At $H=300/360$~Oe in Figs.~\ref{fig1}(a)/(b), the jumps are replaced by large peaks caused by the reversible transition between the P and AP states.~\cite{myprl} The onset of the magnetic dynamics  starting at $I=I_C$ appears as a sharp increase of $dV/dI$ nearly independent of $H$ ($1$~kOe data in Figs.~\ref{fig1}(a),(b)). The approximate equality $I_C\approx I^+$ shown with a dashed line indicates that the reversal occurs when large-amplitude dynamics is excited by ST. The $5$~K data exhibit significantly increased reversal currents and $I_C$.  Fig.~\ref{fig1}(c) summarizes the temperature dependence of $I^+$ and $I^-$. Both are nearly constant above $130$~K, below which they dramatically increase. Similar behaviors of Co/Cu/Co nanopillars indicate their intrinsic origin from the spin-dependent transport in Co.~\cite{CoCo}

\begin{figure}
\includegraphics[scale=0.8]{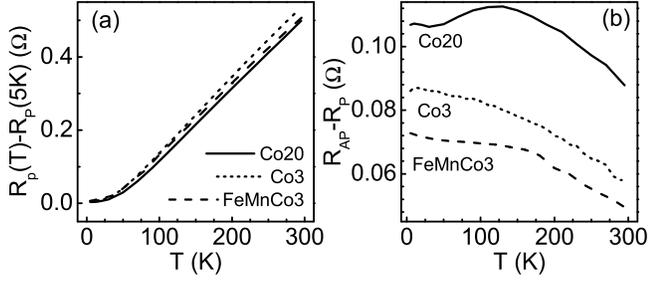}
\caption{\label{fig2} (a) P-state resistances $R_P$ offset by values at $5$~K, and (b) MR {\it vs} $T$ for the three types of samples as labeled.}
\end{figure}

One may attribute some of the dependence on $T$ shown in Fig.~\ref{fig1}(c) to the effects of thermal activation. Indeed, $I^+ \le I_C$ at RT because thermal fluctuations result in reversal slightly before the onset of large-amplitude dynamics. In contrast, $I^+ \ge I_C$ at $5$~K because current-induced dynamics can occur before the reversal occurs. The fundamental quantity predicted by the models of ST is $I_C$. It is insensitive to thermal fluctuations and sample shape imperfections, and can be directly determined from the sharp increase of $dV/dI$ at $H$ large enough to suppress hysteretic reversal.  Fig.~\ref{fig1}(d) summarizes $I_C$ {\it vs} $T$ for all three different sample structures. $FeMnCo3$ data are approximately independent of $T$, while $I_C$ for $Co3$ and $Co20$ increase when $T$ is decreased. Comparing panels (c) and (d) reveals that $I_C$ closely follows $I^+$. It is not possible to measure a similar excitation onset current $I^-_C$ in the AP state, because transition to the P state is not suppressed at any $H$. Below, we use $I^-$ as an approximation for $I^-_C$.

Since F$_2$ is identical in all samples, the different behaviors of $I_C$ must be attributed to the different spin-dependent transport properties of F$_1$. The difference between $Co3$ and $FeMnCo3$ is due to the spin flipping in FeMn, which eliminates spin diffusion in the bottom Cu(50) contact. The difference between the $Co20$ and $Co3$ data indicates that the effects of spin diffusion in Co are stronger than those in Cu. Despite a significant increase of $I_C$ in $Co20$, it does not diverge as would be expected if the sign of ST was reversed.~\cite{fertCoPy} 

Figs.~\ref{fig2}(a),(b) show temperature dependence of $R_P-R_P(0)$ and MR=$R_{AP}-R_P$. $R_P$ increased with $T$ due to magnon and phonon scattering, and were surprisingly consistent among the samples. Interestingly, there is a clear correlation between the variations of MR and $I_C$ in all samples. As $T$ decreases from RT, all MRs increase at a similar rate, while $I_C$ slightly increase. At lower $T$, the trends for $Co3$ and $FeMnCo3$ remain the same, while a decrease of MR in $Co20$ at $T<130$ coincides with a sharp increase of $I_C$.

\begin{figure}
\includegraphics[scale=0.9]{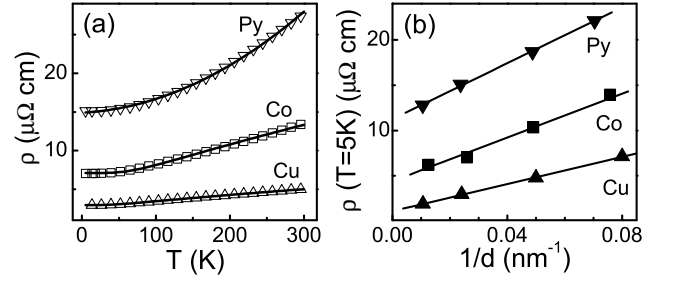}
\caption{\label{fig3} (a) Resistivities of $40$~nm thick Py, Co, and Cu films measured in Van der Pauw geometry. The Co and Cu data are fitted with the Bloch-Gruneisen approximation, with Debye temperatures $\theta_{Co}=373$~K and $\theta_{Cu}=265$~K. The Py data are fitted with a quadratic dependence. (b) Dependencies of residual resistivities on inverse film thickness (symbols), with linear fits shown.}
\end{figure}

To understand the dependencies of MR, CIMS, and $I_C$ on the sample structure, we performed simultaneous calculations of spin-dependent transport and ST. Our model combines a diffusive approximation for the ferromagnets and outer sample contacts with a ballistic approximation for the Cu(10) spacer between the ferromagnets.~\cite{slonczewski2002} This approximation is consistent with calculations based on the Boltzmann equation.~\cite{boltzmann} We combine the continuity conditions for spin currents and spin accumulation in the spacer between F$_1$ and F$_2$ derived by Slonczewski~\cite{slonczewski2002} (Eqs.~(13),(14)) with a small-angle expansion of Eq.~(28) for ST. The resulting expression for ST in terms of the spin current $I_s=I^\uparrow-I^\downarrow$ and spin accumulation $\Delta\mu=\mu^\uparrow-\mu^\downarrow$ in the Cu(10) spacer near the collinear magnetic configuration is
\begin{equation}\label{tau}
\tau=\frac{\hbar\sin(\theta)}{4e}(AG\Delta\mu-I_S)
\end{equation}
where $e$ is the electron charge, $\hbar$ is the Planck's constant, $G$ is twice the mixing conductance introduced in the circuit theory,~\cite{kovalev} $A$ is the area of the nanopillar, and $\theta$ is the angle between the magnetic moments.  At $I=I_C$, $\tau$ compensates the damping torque, yielding
\begin{equation}\label{ic}
I_C=\frac{\alpha e\gamma S_2 2\pi M_2}{\tau},
\end{equation}
where $\alpha\approx 0.03$ is the Gilbert damping parameter,~\cite{cornellscience} $\gamma$ is the gyromagnetic ratio, $\tau$ is ST determined from Equation (\ref{tau}) at $I=1$ in appropriate units, and $S_2=M_2V/2\mu_B$ is the total spin of the Py(5) nanopillar. Here, $V$ is the volume of F$_2$, and $\mu_B$ is the Bohr magneton. The magnetization $M_2$ of Py varied from $730$~emu/cm$^3$ at $20$~K to $675$~emu/cm$^3$ at $300$~K, as determined by magnetometry of Py(5) films prepared under the same conditions as the nanopillars. These values are lower than expected for bulk Py, but consistent with the published results for Py films.~\cite{krivorotovprl}

Equations (\ref{tau}) and (\ref{ic}) express $I_C$ in terms of $\Delta\mu$ and $I_S$, the same quantities that determine MR in magnetic multilayers. We calculated $\Delta\mu$ and $I_S$ self-consistently using a one-dimensional diffusive approximation employing the standard MR parameters: spin asymmetries $\beta$,  renormalized resistivities $\rho^*=\rho/(1-\beta^2)$, spin diffusion lengths $\l_{sf}$ in the layers, and similarly defined parameters $AR^*$, $\gamma$, and $\delta$ for the interfaces.~\cite{valetfert} We estimate these parameters from a combination of the published values~\cite{bassreview} and our own measurements, as described below.

The resistivity of each layer in our samples provides essential information about electron diffusion. Because of variations among published resistivities, we instead determined their values from measurements of thin films prepared under the same conditions as the nanopillars, with thicknesses verified by x-ray reflectometry. Fig.~\ref{fig3}(a) shows $\rho(T)$ for $40$~nm thick Py, Co, and Cu films, together with fittings for Co and Cu with the Bloch-Gruneisen approximation. We obtained better fitting for Py data with a quadratic dependence, indicating that electron-magnon scattering may dominate electron-phonon scattering.~\cite{el-magn} The dependence of the residual resistivity on film thickness was consistent with the Fuchs-Sommerfield approximation (Fig.~\ref{fig3}(b)), allowing us to extract the bulk residual values $\rho_{Py}(0)=11.3$~$\mu\Omega$cm, $\rho_{Co}(0)=4.4$~$\mu\Omega$cm, and $\rho_{Cu}(0)=1.1$~$\mu\Omega$cm. We used the extracted bulk $\rho(T)$ to model all the extended layers in the nanopillars. The effect of lateral confinement in Py(5) nanopillars was approximated by using the resistivity of a Py(40) film.

\begin{figure}
\includegraphics[scale=0.85]{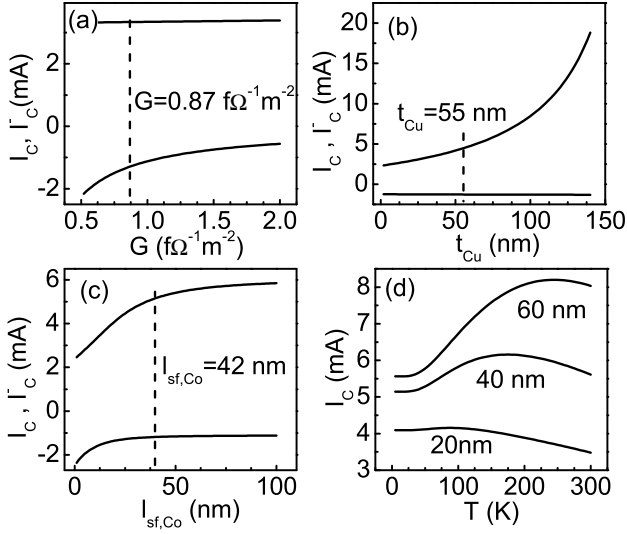}
\caption{\label{fig4} (a) Calculated $I_C$, $I^-_C$ {\it vs} $G$ for $FeMnCo3$. (b) Same {\it vs} $t_{Cu}$ for $Co3$. (c) Same {\it vs} $l_{sf,Co}$ for $Co20$, (d) same {\it vs} $T$ for $Co20$ samples, for the residual values of $l_{sf,Co}$ as labeled.}
\end{figure}

To estimate $l_{sf}(T)$, we used its empirical inverse relationship with $\rho$, along with the bulk residual values $l_{sf, Py}(0)=6$~nm, and $l_{sf, Cu}(0)=300$~nm based on published measurements,~\cite{bassreview} scaled by the somewhat different residual resistivities of our films. If scattering by thermal excitations does not flip electron spins, a weaker dependence $l_{sf}(T)\propto\sqrt{1/\rho(T)}$ is possible. However, we show below that a dependence even stronger than $1/\rho$ is more likely. We use $\beta_{Py}=\gamma_{Py/Cu}=0.7$, $\gamma_{Co/Cu}=0.8$, $\beta_{Co}=0.36$ for spin asymmetries,   $AR^*_{Co/Cu}=0.55$~$f\Omega m^2$, $AR^*_{Py/Cu}=0.5$~$f\Omega m^2$ for renormalized interface resistances, and $\delta_{Co/Cu}=0.2$, $\delta_{Py/Cu}=0.25$ for spin flipping coefficients.~\cite{CoCo,bassreview} Their dependence on $T$ is neglected due to the dominance of the band structure and impurity scattering far from the Curie temperature. For FeMn, we used $l_{sf,FeMn}\approx 0.5$~nm, and $\rho_{FeMn}=87$~$\mu\Omega$cm. Scattering at its interfaces was modeled by adding $0.5$~nm to the nominal thickness of FeMn.  To account for the Cu contacts, the calculation included outer Cu layers of thickness $t_{Cu}$, determined as described below. These layers were terminated with fictitious spin sinks. 

To demonstrate that CIMS is extremely sensitive to the effects of diffusion, we now describe how our $5$~K data can be fitted by appropriate choice of three parameters whose values have the largest uncertainty: conductance $G$ in Equation (\ref{tau}), effective MR-active thickness $t_{Cu}$ of the Cu contacts, and spin diffusion length $l_{sf, Co}$. Calculations for $FeMnCo3$ were significantly affected only by $G$, which controls the asymmetry of CIMS. The values of $I_C/|I^-|$ in mA measured at $5$~K for three $FeMnCo3$ samples were $2.3/0.8$, $1.6/0.6$, and $3.1/1.5$, giving an average ratio $I_C/|I^-|=2.6$.   The calculated value increases from $1.46$ at  $G=0.5$~$f\Omega^{-1} m^{-2}$ to $6.1$ at $G=2$~$f\Omega^{-1} m^{-2}$ (Fig.~\ref{fig4}(a)). The best values $I_C/|I^-_C|=3.34/1.27$ are obtained at $G=0.87$~$f\Omega^{-1} m^{-2}$, in reasonable agreement with band structure calculations.~\cite{slonczewski2002,xia} 

Spin diffusion in the bottom Cu layer has little effect on $Co20$ and $FeMnCo$ due to the spin relaxation in Co and FeMn, respectively.  To determine $t_{Cu}$, we use the ratios $I_C/|I^-|$ of the three $Co3$ samples, $3.55/1.0$, $4.6/1.5$, and $4.2/1.2$, giving an average ratio $I_C/|I^-|=3.4$. The calculated $I_C/|I^-_C|$ increases from $1.9$ for $t_{Cu}=0$ to $14$ for  $t_{Cu}=140$~nm (Fig.~\ref{fig4}(b)), and eventually diverges at $t_{Cu}=200$~nm. The best agreement with data is obtained for $t_{Cu}=55$~nm, resulting in $I_C/|I^-_C|=4.4/1.3$. 

\begin{figure}
\includegraphics[scale=0.85]{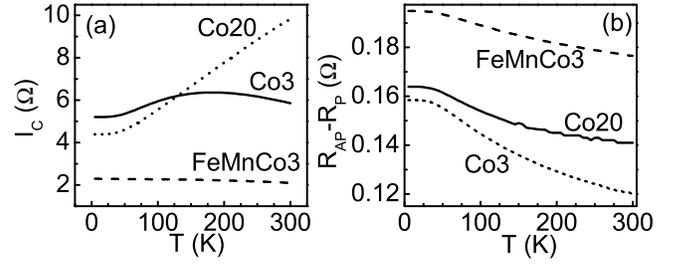}
\caption{\label{fig5} (a) Calculated $I_C$ vs $T$, and (b) calculated MR {\it vs} $T$ for three sample types as labeled.}
\end{figure}

Lastly, diffusion in Co significantly affects CIMS in samples $Co20$, but not in $Co3$ and $FeMnCo3$. We determine $l_{sf, Co}$ from the ratio $I_C/|I^-|$ of five $Co20$ samples, $8.9/2.1$, $7.3/1.6$, $9.0/2.0$, $8.5/2.0$, $8.0/1.7$, giving an average ratio $I_C/|I^-|=4.4$. Fig.~\ref{fig4}(c) illustrates that the calculated ratio $I_C/|I^-_C|$ increases from $1.0$ for $l_{sf,Co}=0$ to $5.2$ for $l_{sf,Co}=100$~nm. The best agreement with the data is obtained for $l_{sf,Co}=42$~nm consistent with the published values.~\cite{bassreview}  

Despite the ability to model the $5$~K data, the calculations did not reproduce the dramatic dependence of $I_C$ on $T$ in Fig.~\ref{fig2} (see below). Therefore, one can attempt to determine $l_{sf,Co}$
from the dependence of $I_C$ on $T$. Fig.~\ref{fig4}(d) shows calculations for the residual values $l_{sf,Co}=20$~nm, $40$~nm, and $60$~nm. Large $l_{sf,Co}$ results in $I_C$ decreasing with $T$, which is inconsistent with the data. Small $l_{sf,Co}$ gives decrease of $I_C$ with $T$ in better qualitative agreement with data, but gives unreasonably small $I_C$ at $5$~K. Consequently, we return to the value determined from Fig.~\ref{fig4}(c).

Fig.~\ref{fig5}(a) shows the calculated $I_C$ {\it vs} $T$ for the three sample types. To interpret these results, we note that Figs.~\ref{fig4}(b)-(d) exhibited an increase of $I_C$ when the effective MR-active resistance of F$_1$ determined by $\rho l_{sf}$ was increased. This relationship was also established analytically.~\cite{slonczewski2002,myagmr} The experimental correlation between the decreases of MR and increases of $I_C$ in Figs.\ref{fig1},~\ref{fig2} is of the same origin.  The lack of temperature dependence for $FeMnCo3$ is therefore consistent with negligible spin diffusion effects in F$_1$. In calculations for $Co3$, the increase of $I_C$ with $T$ is caused by the increased contribution $t_{Cu}l_{sf,Cu}$ to the effective resistance of F$_1$. Calculations for $Co20$ show a competition between the contribution of the bulk Co resistivity, which increases with $T$, and the contributions from the Cu(50) layer and the outer Co/Cu interface, which decrease with $T$ due to the increased spin flipping in Co. However, both $Co3$ and $Co20$ calculations do not reproduce the data, suggesting that the effects of thermal scattering should be re-examined. 

The calculated dependence of MR on $T$ was in overall agreement with data for $Co3$ and $FeMnCo3$, but did not reproduce the decrease at $T<130$~K seen in $Co20$ data (Fig.~\ref{fig5}(b)). The calculations overestimated the values, suggesting that our samples may be larger than their nominal size. However, this seems to contradict the calculated temperature dependence of $R_P$ consistent with the data (not shown), and the values of $I_C$ that are {\it larger} than the measured $5$~K values. This discrepancy can be reduced e.g. by decreasing $l_{sf,Py}$, which results in a decreased MR without significantly affecting CIMS. 

The failure of $Co20$ calculations to capture the decrease of MR and the increase of $I_C$ at $T<130$~K indicates that $l_{sf,Co}$ decreases with $T$ more rapidly than the accepted $l_sf\propto 1/\rho$, resulting in the reduction of the effective MR-active resistance $\rho_{Co}l_{sf,Co}$. One possible mechanism for such a strong dependence may be electron-magnon scattering which can result in electron spin flipping without significant momentum scattering.  We leave more detailed and perhaps alternative explanations to future studies.

To summarize, we performed magnetoelectronic measurements of nanopillars with three different  polarizing magnetic layers. The samples exhibited different current-induced behaviors, attributed to the spin diffusion in the polarizing layer. The calculations reproduced the lower temperature behaviors with reasonable values of transport parameters. However, temperature dependencies of magnetoresistance and current-induced switching indicate that the effects of thermal scattering on spin transport are more significant than presently believed.

We thank Mark Stiles, Jack Bass and Norman Birge for helpful discussions. This work was supported by NSF Grant DMR-0747609 and a Research Corporation Cottrell Scholar Award. SB acknowledges support from the NASA Space Grant Consortium.

\end{document}